%% file: main.tex
\documentclass[a4paper, amsfonts, amssymb, amsmath, reprint, showkeys, twoside, superscriptaddress,amsart]{revtex4-1}
\usepackage[english]{babel}
\usepackage{mathrsfs}
\usepackage{amsmath,amssymb,graphicx}
\usepackage{nccmath}
    {\end{pmatrix}\end{medsize}}%
\usepackage[utf8]{inputenc}
\usepackage[colorinlistoftodos, color=green!40, prependcaption]{todonotes}
\input{preamble}
\usepackage[pdftex, pdftitle={Article}, pdfauthor={Author}, colorlinks=true, linkcolor=blue, citecolor=blue, urlcolor=blue]{hyperref}
\usepackage{color}
\usepackage{mathtools}
\usepackage{bbm}

\usepackage{float}
\usepackage{array}
\usepackage{longtable}

\usepackage{verbatim}

\usepackage[utf8]{inputenc}
\usepackage[T1]{fontenc}
\usepackage{etoolbox}

\usepackage{anyfontsize}  
\usepackage{microtype} 
\usepackage{newtxtext}
\usepackage{newtxmath}
\usepackage[normalem]{ulem}

\newcommand{\overbar}[1]{\mkern 1.5mu\overline{\mkern-1.5mu#1\mkern-1.5mu}\mkern 1.5mu}
\begin{document}

\title{Polarization-Entanglement Dynamics in Optical Fibers: Mitigating Decay in the Non-Markovian Regime with Dynamical Decoupling}

\author{Pratik J. Barge\footnote{author footnote}}
\email[]{pbarge1@lsu.edu}
\affiliation{Hearne Institute for Theoretical Physics, Department of Physics \\
and Astronomy, Louisiana State University, Baton Rouge, Louisiana 70803, USA.}
\author{Arshag Danageozian}
\affiliation{Department of Physics, Virginia Tech, and Virginia Tech Center for Quantum Information Science and Engineering, Blacksburg, Virginia 24061, USA}
\author{Manish K. Gupta\footnote{author footnote}}
\email[]{Initial investigation was proposed and done by MKG}
\affiliation{AstroCeNT --- Particle Astrophysics Science and Technology Centre --- 
International Research Agenda, \\
Nicolaus Copernicus Astronomical Center, Polish Academy of Sciences,\\
Rektorska 4, 00-614 Warsaw, Poland}
\author{\\Brian T. Kirby\footnote{author footnote}}
\email[]{brian.t.kirby4.civ@army.mil}
\affiliation{DEVCOM Army Research Laboratory, Adelphi, MD 20783, USA}
\affiliation{Tulane University, New Orleans, LA 70118, USA}
\author{Hwang Lee\footnote{author footnote}}
\email[]{hwlee@lsu.edu}
\affiliation{Hearne Institute for Theoretical Physics, Department of Physics \\
and Astronomy, Louisiana State University, Baton Rouge, Louisiana 70803, USA.}

\date{\today}

\begin{abstract}
Future distributed quantum systems and networks are likely to rely, at least in part, on the existing fiber infrastructure for entanglement distribution; hence, a precise understanding of the adverse effects of imperfections in optical fibers on entanglement is essential to their operation. Here, we consider maximally entangled polarization qubits and study the decay of the entanglement caused by spatial fluctuations in the refractive index of optical fibers. We study this entanglement dynamics using the spin-boson model and numerically solve our system of equations using the hierarchical equations of motion (HEOM) formalism. We show that within the range of practically relevant system parameters, our developed model exhibits both Markovian and non-Markovian entanglement decay behavior. Further, to counter the observed entanglement decay, we propose the implementation of dynamical decoupling in optical fibers using spaced half waveplates. In particular, we numerically model the time-dependent Hamiltonians of the Carr-Purcell-Meiboom-Gill and Uhrig dynamical decoupling schemes and show a reduced rate of entanglement decay even with sparsely spaced half waveplates along the length of optical fiber. Finally, we evaluate the performance of these two schemes in multiple system configurations.
\end{abstract}

\maketitle

\section{Introduction}
The distribution of quantum entanglement to separate parties over long distances is a prerequisite for the existence of quantum technologies such as quantum key distribution \cite{bennett2020quantum,PhysRevLett.67.661}, distributed quantum computation \cite{gottesman1999demonstrating, cirac1999distributed}, distributed quantum sensing \cite{komar2014quantum, gottesman2012longer} and the quantum internet \cite{kimbleQNet}. Photons are widely considered to be the most suitable carriers of quantum information for long-distance applications due to their long coherence times, limited environmental interactions and ability to leverage existing fiber optical distribution network. Optical quantum information processing platforms exploit quantum correlations encoded in different degrees of freedom of photons including path, number, polarization and frequency. Among these, the polarization-encoding, where the two states of a qubit are encoded in the two orthogonal polarization states of a photon, is particularly appealing due to its ease of generation, manipulation, and detection using standard optical components.
\begin{figure*}[ht]
    \centering
    \includegraphics[width=\textwidth]{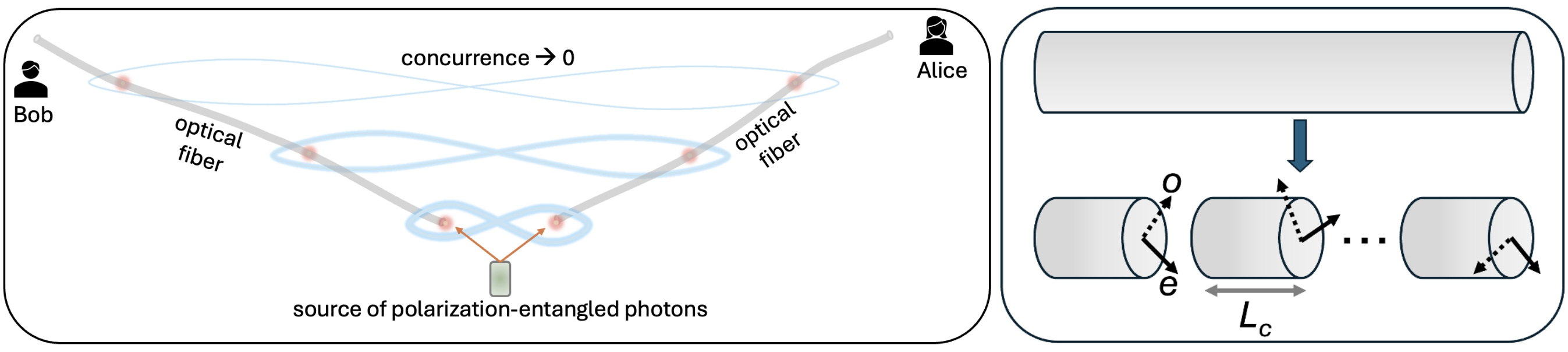}
    \renewcommand{\figurename}{Figure}
    \caption {Schematic of the problem: (left) distribution of polarization entanglement to Alice and Bob via optical fibers: concurrence decays as maximally entangled two-photon Bell state propagates through two long, birefringent single-mode optical fibers.  (right) such a fiber can be modeled as a series of segments, each with a random birefringence magnitude and a random orientation of the ordinary (o) and extraordinary (e) axes.} 
    \label{fig:problem_schematic}
\end{figure*}
Practical optical fibers have imperfections that cause decoherence in polarization qubits. These arise from shape asymmetries due to manufacturing or external factors like stress and temperature changes. This results in birefringence, where different polarizations experience different refractive indices. The birefringence fluctuates randomly in magnitude and direction along the fiber, causing orthogonal polarization modes to travel at different velocities, leading to polarization mode dispersion (PMD)~\cite{nelson2004introduction, hui2022fiber}. Although polarization-entangled photons have been transmitted over $100$ km in free space \cite{ursin2007entanglement}, the presence of birefringent effects impede long-distance transmission of quantum information through optical fibers. 
Birefringence has been previously studied considering loss of coherence and entanglement~\cite{antonelli2011sudden, lim2016effects, brodsky2011loss, salemian2011analysis, jones2020exploring, huang2007fiber} as well as other effects~\cite{liu2023quantitative, yaman2004effects} in fiber-based systems.

Numerical and experimental studies~\cite{wengerowsky2019entanglement, shen2022distributing,  hubel2007high, lee2024non, salemian2011analysis} have reported propagation distances from a few km to more than 100 km for transmitting polarization-encoded qubits via fiber, varying with system configurations (e.g. number of optical paths exposed to PMD effects, compensation mechanisms) and operating conditions (e.g. fiber quality, pump bandwidth, transmission rate).
In classical communication systems repeaters are used to increase the transmission distance by amplifying and resending the signal. Quantum communication systems operate within the confines of the no-cloning theorem and have to employ quantum repeaters that use a pair of entangled states and Bell measurement to perform entanglement swapping~\cite{noClone82, PhysRevLett.81.5932}. However, increasing the number of repeaters leads to increased network complexity, losses, and the need to find efficient strategies to distribute entanglement in quantum repeater networks~\cite{daSilva2021genetic,Khatri2021policieselementary,li2020efficient,haldar2024fast}. 

The study of open quantum systems (OQS) reveals how interactions with external environments affect quantum systems and aims to control and mitigate unwanted effects. 
For system-environment interactions characterized by weak coupling and short environmental correlation times, the quantum system's evolution can be approximated as Markovian and described by the Lindblad master equation; however, when the environment exhibits memory effects, the dynamics become non-Markovian~\cite{breuer2002theory,schlosshauer2014quantum}. Birefringence-induced decoherence effects have been relatively underexplored from the OQS perspective. While a master equation with Markovian assumption is developed in Ref~\cite{poon2008polarization}, non-Markovian effects have been shown to arise due to factors such as frequency correlated signal and idler photons~\cite{liu2013photonic}, strong coupling and loss~\cite{lee2024non}.

In this work we first develop a model for the decay of polarization entanglement of Bell state photons propagating through single-mode fibers, considering possible non-Markovian effects emerging due to long correlation length and large variance in birefringence of fiber. For this purpose, we use hierarchical equations of motion (HEOM)~\cite{tanimura1989time}, one of the limited number of techniques for calculating non-Markovian dynamics of OQS. Unlike frequently used Lindblad master equation, these are derived without making perturbative Born-Markov or secular approximations. After identifying distinct disentanglement behaviours in different parameter regimes we then numerically simulate the use of dynamical decoupling (DD) in the optical fibers for preserving the entanglement. Ref.~\cite{PhysRevA.70.062310} investigated the general noise properties of fiber channels and first proposed application of DD in optical fibers for protecting photonic qubits. It has been studied in the context of a single qubit~\cite{PhysRevA.85.022340,PhysRevA.91.032329} and orbital angular momentum qudits~\cite{Gupta1, gupta2016minimizing} in optical fibers. In contrast, this work uses HEOM formalism to focus on polarization-entanglement dynamics in optical fibers in Markovian and non-Markovian regimes and to evaluate the effectiveness of dynamical decoupling schemes.

The organization of the paper is as follows:~In Section~\ref{sec:theory}, we describe our theoretical model for birefringence-induced loss of entanglement and briefly review the HEOM formalism. In Section~\ref{sec:decay-noDD}, we first map the decoherence of the entangled bipartite state propagating in the fibers onto the spin-boson model and simulate it using the HEOM formalism. Finally in section~\ref{sec:DD}, we simulate and assess the impact of two DD schemes on preserving entanglement in single-mode fibers.

\section{Theoretical model}
\label{sec:theory}
In this section, we present our model of birefringence-induced decoherence for polarization-entangled Bell states traveling in optical fibers. A family of two-photon polarization-entangled state can be represented in the computational basis $\{\ket{HH},\ket{HV},\ket{VH},\ket{VV}\}$ as:
\begin{equation}
\ket{\Psi^{\pm}_{\textrm{in}}}=\frac{1}{\sqrt{2}} (\ket{HV} \pm \ket{VH} ),
\end{equation}
\begin{equation}
\ket{\Phi^{\pm}_{\textrm{in}}}= \frac{1}{\sqrt{2}} (\ket{HH} \pm \ket{VV}).
\end{equation}
Concurrence is one of the widely used measures to quantify the entanglement of a bipartite quantum state. For a quantum state $\rho$ it is given by:
\begin{align}
C\left(\rho\right)= \max \lbrace  0, \lambda_{1}-\lambda_{2}-\lambda_{3}-\lambda_{4} \rbrace \,
\end{align}
where $\lambda_i$ are the square roots of the eigenvalues of the matrix $R^2 = \rho \cdot \Tilde{\rho}$. Here $\Tilde{\rho} = (\sigma_y \otimes \sigma_y) \rho^* (\sigma_y \otimes \sigma_y) $ and $\sigma_y$ is Pauli matrix.

The decoherence of the polarization qubit in optical fibers has its origin in birefringence ($\Delta \beta$). A single-mode fiber supports two degenerate orthogonal modes due to its circular symmetry. But real fibers have some asymmetry due to manufacturing imperfections or external stresses such as bends and twists on the fiber due to deployment in the field. This breaks the degeneracy of two orthogonal modes and they experience different effective refractive indices. This results in birefringence, difference between the propagation constants of both modes:
\begin{equation}
\Delta \beta = \frac{2\pi}{\lambda}\Delta n = \frac{\omega}{c}\Delta n,
\end{equation}
where $\Delta n = |n_o - n_e|$ is the difference in refractive index of ordinary (o) and extraordinary (e) modes, $\omega$ is the angular frequency, $c$ is the speed of light and $\lambda$ is the wavelength. For a polarization qubit with initial state 
\begin{equation}
\ket{\psi (0)} = \frac{1}{\sqrt{2}}[\ket{H} + \ket{V}],
\end{equation}
after propagating for a length $L$ in a medium of constant birefringence evolves to 
\begin{equation}
\ket{\psi (t)} = \frac{1}{\sqrt{2}}[\ket{H} + e^{-i\phi}\ket{V}].
\end{equation}
The accumulated phase is given by 
\begin{equation}
\phi = \Delta \beta l = \frac{2\pi L}{\lambda} \Delta n.
\label{eq:qubit-phase}
\end{equation}

For a given optical fiber, the birefringence remains constant for a length known as the "correlation length" ($L_c$). Practical fibers are long, and variation of birefringence along its length is stochastic in nature. As shown in Fig.\ref{fig:problem_schematic} (right), such a fiber can be viewed as a series of sections, each of length $L_c$, where both $\Delta n$ and the alignment of the fast and slow axes vary randomly from section to section. This means the birefringence axes rotate randomly along the fiber. This causes randomization of the accumulated phase and leads to decay of coherence as well as entanglement.

We analyze the decoherence effects in birefringent media using the theory of open quantum systems. We model the decoherence of the photonic qubits due to stochastic fluctuations in birefringence through an open system losing coherence after interacting with environmental degrees of freedom. The system of interest is a photonic qubit, and birefringence noise is modeled as its interaction with a bath of harmonic oscillators. The dynamics of this process are generally governed by the nature of the fluctuations. The evolution of the system's density matrix is determined without directly specifying the bath degrees of freedom, but rather by incorporating the impact of the bath on it via the bath spectral density or, equivalently, via the bath correlation function.

\subsection{Hierarchical Equations of Motion} 
We begin this section by reviewing the HEOM formalism that enables tracking the time evolution of a system's reduced density matrix while capturing non-Markovian effects of system-bath dynamics through \textit{auxiliary} density matrices (ADM). The Lindblad equation, derived under the Born-Markov and secular approximation, cannot capture non-Markovian behavior. Consider the spin-boson model, a commonly used phenomenological quantum model, where interaction with a quantized bath of harmonic oscillators provides stochastic fluctuations. The dynamics of a spin-$\frac{1}{2}$ particle interacting with such a bath are governed by the Hamiltonian
\begin{equation}
H = H_S + H_B + H_{SB},
\end{equation}
where $H_S$ governs free evolution of the system. Hamiltonians for the bath and the system-bath interaction are
\begin{equation}
H_B + H_{SB} = \sum_{k} \hbar \omega_k \hat{b}_k^\dagger \hat{b}_k + \hat{Q} \sum_{k} \hbar g_k(\hat{b}_k + \hat{b}^\dagger_k),
\label{eq:bath-int}
\end{equation}
where $\hat{b}^\dagger_k$ and $\hat{b}_k$ are the creation and annihilation operators associated with the $k$th mode of the bath, $\omega_s$ and $\omega_k$ are frequencies of the system and kth bath mode, and $\hat{Q}$ is the system operator that couples, with coupling strengths $g_k$, to the bath operators.

System-bath interaction and the effect of the bath on the system are described through the spectral density of the oscillator modes of the bath
\begin{equation}
J(\omega) = \sum_{k} \left | g_k \right |^2 \delta(\omega - \omega_k).
\end{equation}
A Gaussian bath can be fully specified by its second-order correlation function and is related to the spectral density of the bath modes as
\begin{equation} 
\begin{split}
\mathcal{C}(\tau) = &\expval{\mathcal{U}(t+\tau) \mathcal{U}(t)}  \\
  = &\frac{1}{\pi}\int_{0 }^{\infty}d\omega J(\omega) \frac{e^{i\omega\tau}}{1-e^{\beta\hbar\omega}}
\end{split}
\end{equation}
with collective bath coordinate $\mathcal{U} = \sum_{k} g_k (\hat{b}_k + \hat{b}^\dagger_k)$ and bath temperature $T = 1/\beta$.

For certain $J(\omega)$ e.g. Debye or Lorentzian, corresponding $\mathcal{C}(t)$ can be expanded as sum of exponentials using Matsubara decomposition \cite{liu2014reduced}:
\begin{equation}
\mathcal{C}(t) = \mathcal{C}_R(t) + i\mathcal{C}_I(t) = \sum_{m=0}^{M_R} c_m^R e^{-v_m^R t}+\sum_{m=0}^{M_I} c_m^I e^{-v_m^I t},
\label{eq:matsubara}
\end{equation}
where the subscripts $R$ and $I$ denote the real and imaginary parts of correlation function, $v_m$ are bosonic Matsubara frequencies and $c_m$ are coefficients with $\{v_m, c_m\} \in \mathbb{C}$.
The HEOM, as derived in Ref.~\cite{tanimura1989time, strumpfer2009light, ma2012entanglement, lambert2023qutip} by using the correlation function expansion of Eq.~(\ref{eq:matsubara}), is a set of coupled, ordinary differential equations of the form
\begin{align}
    \frac{d \rho_\textbf{n} (t)}{dt} = \hspace{0.5em} & \Bigl ( -i H_S^\times -\sum_{j=R,I}^{} \sum_{k=1}^{M_j} n_{jk} v_{jk} \Bigr )\rho_\textbf{n}(t) \nonumber \\ 
    & - i \sum_{k=1}^{M_R} c_k^R n_{Rk} \hat{Q}^\times \rho_{n^-_{Rk}}(t) + \sum_{k=1}^{M_I} c_k^I n_{Ik} \hat{Q}^\circ \rho_{n^-_{Rk}}(t) \nonumber \\
    & + \sum_{j=R,I}^{} \sum_{k=1}^{M_j} \hat{Q}^\times  \rho^+_{jk}(t)
   \label{eq:heom_set}
   \end{align}
where $\hat{Q}$ is the system operator that couples to the environment. Density matrices are labeled by vector index $\textbf{n} = \{n_{R1}, n_{R2}..n_{RM_R}, n_{I1}, n_{I2}...n_{IM_I}\}$ with integers $n_{jk}\in \{0..\infty\}$, $M$ being number of terms in Matsubara expansion and $\rho_{\textbf{n}^{\pm}_{jk}}$ are density matrices at hierarchy levels higher/lower than those indexed by $\rho_{\textbf{n}}$, obtained by raising or lowering the index $n_{jk}$ by one. 
Theoretically, the value of integers $n_{jk} \in \{0,..\infty\}$, in practice it is limited up to a truncation level $N_c$ to make HEOM numerically feasible. Required value of $N_c$ scales with level of non-Markovianity of the dynamics and is set by considering the convergence of numerical solution. RDM of the system is stored in the density matrix labeled with $\textbf{n} = 0 = \{0,0,..0\}$ and belongs to hierarchy level zero. Note that the ADMs with $\textbf{n} \neq 0$ are not physical states but mathematical constructs that capture non-Markovian effects of the system-bath interaction.

\begin{figure}[h]
    \centering
    \includegraphics[width=\columnwidth]{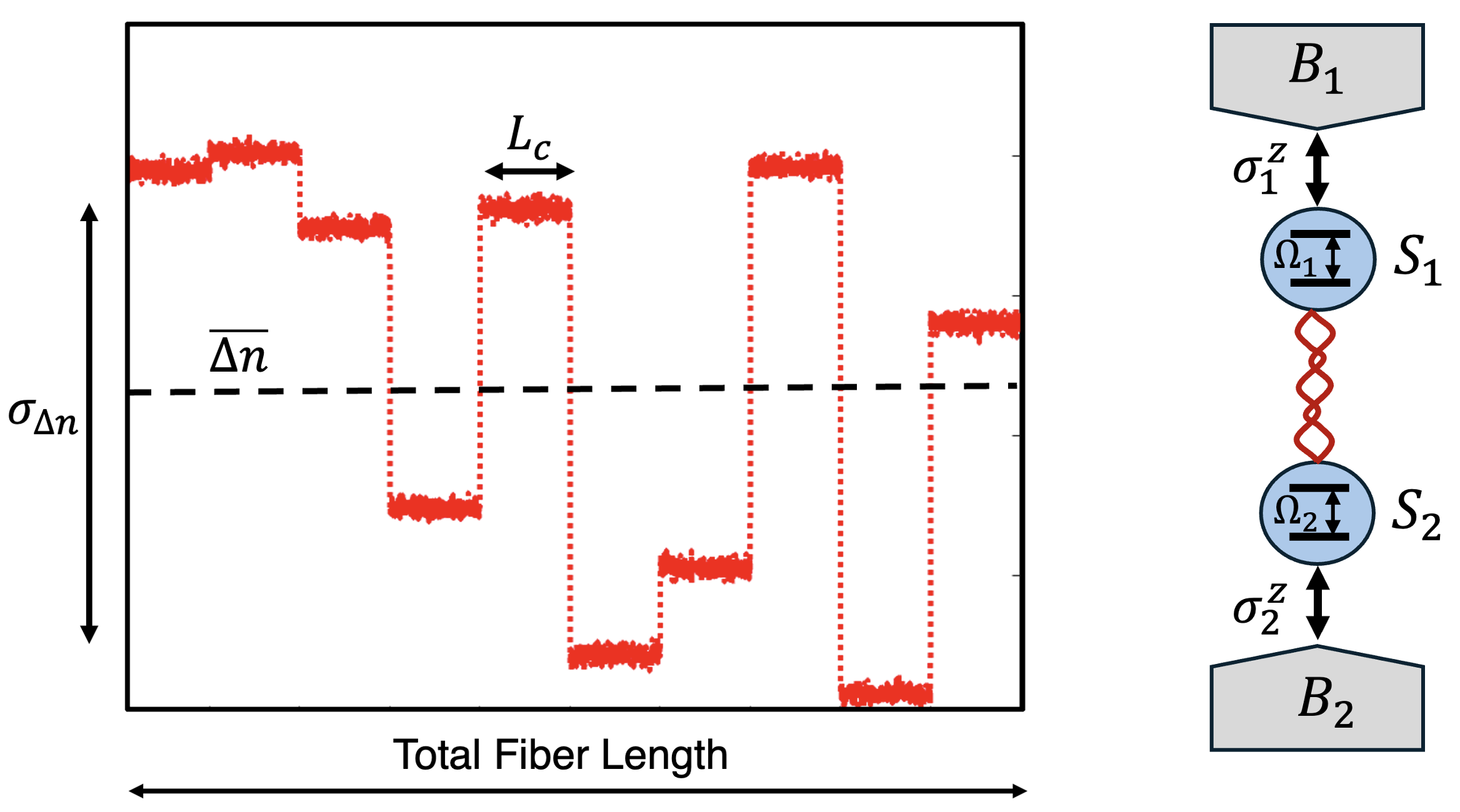}
    \renewcommand{\figurename}{Figure}
    \caption {(Left) sample realization of fluctuations in birefringence of fiber with mean birefringence $\overbar{\Delta n}$,  standard deviation $\sigma_{\Delta n}$ and correlation length $L_c$. (Right) schematic of the model: interaction between qubit system $S_{i}$ and bath $B_{i}$ is coupled by system operator $\sigma_i^z$.}
    \label{fig:model}
\end{figure}
\section{Birefringence-induced Entanglement decay}
\label{sec:decay-noDD}
In this section, we employ the spin-boson model and the HEOM formalism to investigate the entanglement dynamics of two polarization-entangled photons traveling in optical fiber. Each photon propagates through a distinct single-mode optical fiber, where the fibers, though different in realization, possess identical average characteristics such as length and birefringence values.
The decoherence effects of stochastic fluctuations in fiber birefringence are modeled as the interaction of two entangled two-level systems (TLS) $S_{1,2}$ with a bath of harmonic oscillators $B_{1,2}$ as shown in Figure \ref{fig:model}. We draw parallels between parameters of this phenomenological model and physical properties of single-mode fibers in the following way: beat length $L_b=\lambda/\overbar{\Delta n}$ is the distance over which the polarization state rotates by $2\pi$ in a fiber with average birefringence $\overbar{\Delta n}$. Using eq.~\ref{eq:qubit-phase} it can be connected to the frequency of the TLS as follows: since accumulated phase $\phi = \frac{2\pi L}{\lambda} \overbar{\Delta n} = \omega \cdot \overbar{\Delta n} \cdot t $, the qubit frquency $\Omega = \frac{d\phi}{dt} = \omega \cdot \overbar{\Delta n}$.
Bath correlation time $\tau_{c}$ can be associated with the correlation length $L_{c}$ as
\begin{equation}
    \tau_c = \frac{L_c}{v_f} = \frac{L_c}{c}n_f 
\end{equation}
where $v_f = c/n_f$ is the speed of light in the fiber, $n_f$ is the average refractive index of the fiber. We parameterise the strength of the coupling between the qubit and the birefringence noise bath in terms of the magnitude of the TLS frequency fluctuations ($\delta\Omega$) as
\begin{equation}
   \eta = \frac{\delta\Omega}{\Omega} = \frac{\sigma_{\Delta n}}{\overbar{\Delta n}},
\end{equation}
where $\sigma_{\Delta n}$ is the standard deviation of fluctuations in $\Delta n(x)$.

We consider the following system Hamiltonian
\begin{equation}
    \label{eq:hamiltonian}
    H_S = \frac{\hbar}{2} \left(\Omega_1 \hat{\sigma}_{1}^z \otimes \hat{\mathbb{I}}_2 + \hat{\mathbb{I}}_1 \otimes \Omega_2 \hat{\sigma}_{2}^z  \right )
\end{equation}
where $\hat{\sigma}_{1,2}^z$ are the Pauli Z operators and $\Omega_{1,2}$ are frequencies associated with $L_b$'s of two fibers. For simplicity we assume both fibers have identical properties, on average, so that $\Omega_{1}=\Omega_{2}=\Omega_{0}$. $H_B$ and $H_{SB}$ are as specified in Eq.~\ref{eq:bath-int} with coupling operator 
\begin{equation}
    \hat{Q} = \hat{\sigma}_{1}^z \otimes \hat{\mathbb{I}}_2 + \hat{\mathbb{I}}_1 \otimes \hat{\sigma}_{2}^z.
\end{equation}
Two entangled TLS interact with two separate but identical baths via this coupling operator. Fig.~\ref{fig:model} (left) shows a realization of random fluctuations of fiber birefringence around the mean value $\overbar{\Delta n}$ throughout the total fiber length. This motivates the choice of bath correlation function,
\begin{equation}
\label{eq:correlation-function}
    C(t_1 - t_2) = \eta \exp\bigl(\frac{-v_f| t_2 - t_1|}{L_c} \bigr),
\end{equation}
in our model where noise correlation decays over length scale of $L_c$.

We use QuTiP's HEOM library \cite{johansson2012qutip, lambert2023qutip} to numerically evaluate Eq.~\ref{eq:heom_set}. The correlation function in Eq.(\ref{eq:correlation-function}) is readily in the required Matsubara form with $M = 1$ and with real $(R)$ and imaginary $(I)$ coefficients
$c^R = \eta$,  $v^R =\gamma$ and $c^I = v^I = 0$ 
in the expansion of $C(t) = c^R e^{-v^R t} + c^I e^{-v^I t}$. In our phenomenological model, we use the following values of the single-mode fiber parameters: typically $\overbar{\Delta n}\approx 10^{-5}-10^{-7}$ and with $\lambda=1550$ nm, $L_b \approx 10$ m \cite{hui2022fiber, salemian2011analysis}. The correlation length $L_c$ can vary from $1$ m to $1$ km depending upon manufacturing and operating conditions \cite{nelson2004introduction} which consequently dictates the range of values for correlation time $1/\gamma$. We set $\overbar{\Delta n} = 10^{-7}$, $\omega \approx 10^8 \mu s^{-1}$, $n_f = 1.5$, $v_f = 0.2$ km/$\mu$s and hierarchy truncation at $N_c = 10$.
\begin{figure}[h]
    \centering
    \includegraphics[width=0.9\columnwidth]{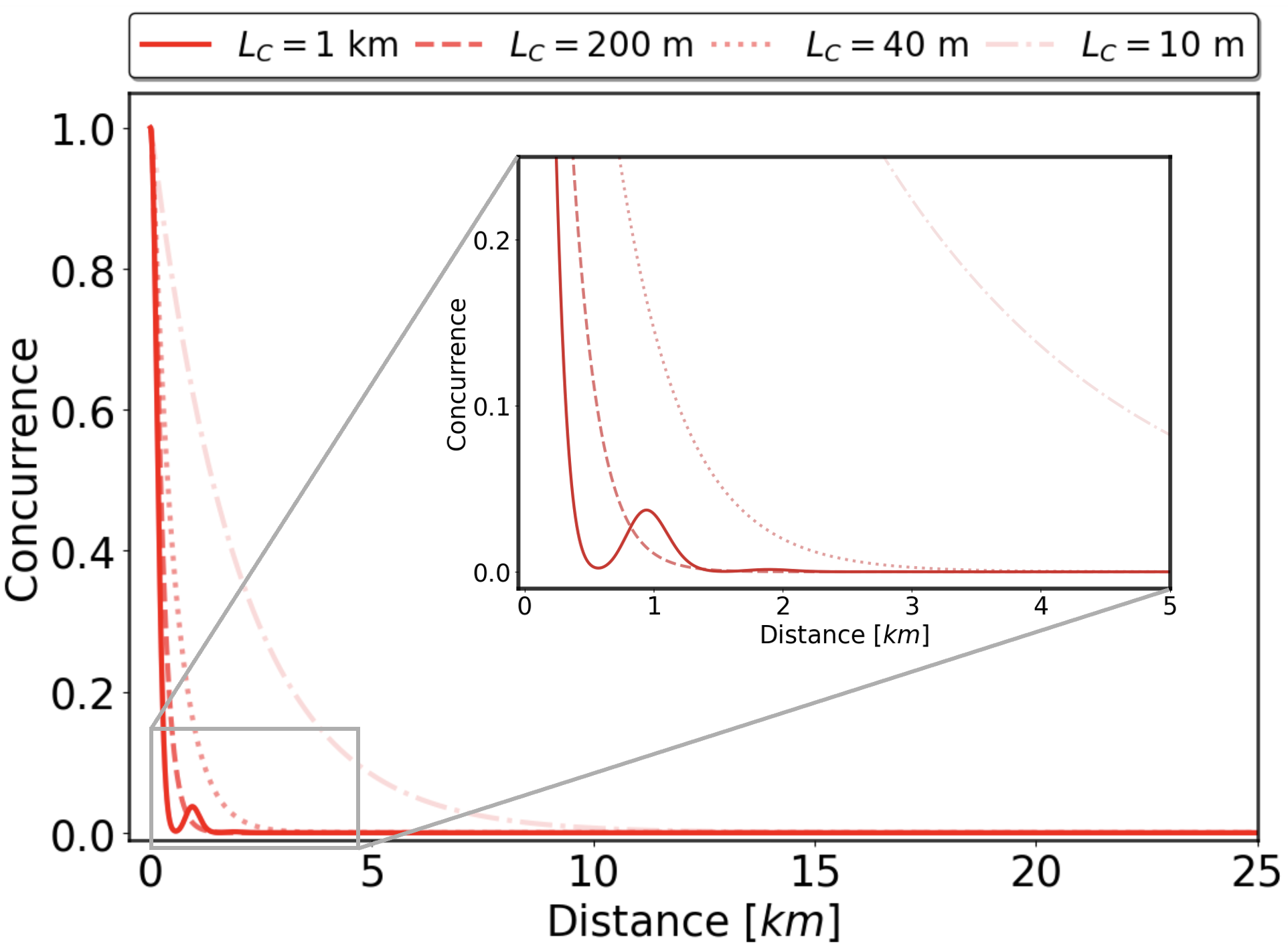}
    \renewcommand{\figurename}{Figure}
    \caption {entanglement decay with propagation distance for fixed coupling strength $\eta = 0.1$ and different correlation lengths $L_c$ = $\{$10 m, 40 m, 200 m, 1 km $\}$. Entanglement decays relatively slowly and monotonically when bath correlation time is lower. A closer look at the early evolution of the entanglement and a tiny revival of the entanglement in the red curve can be seen in the inset.}
    \label{fig:concur_vs_distance_noDD_gammas}
\end{figure}

\begin{figure}[h]
    \centering
    \includegraphics[width=0.9\columnwidth]{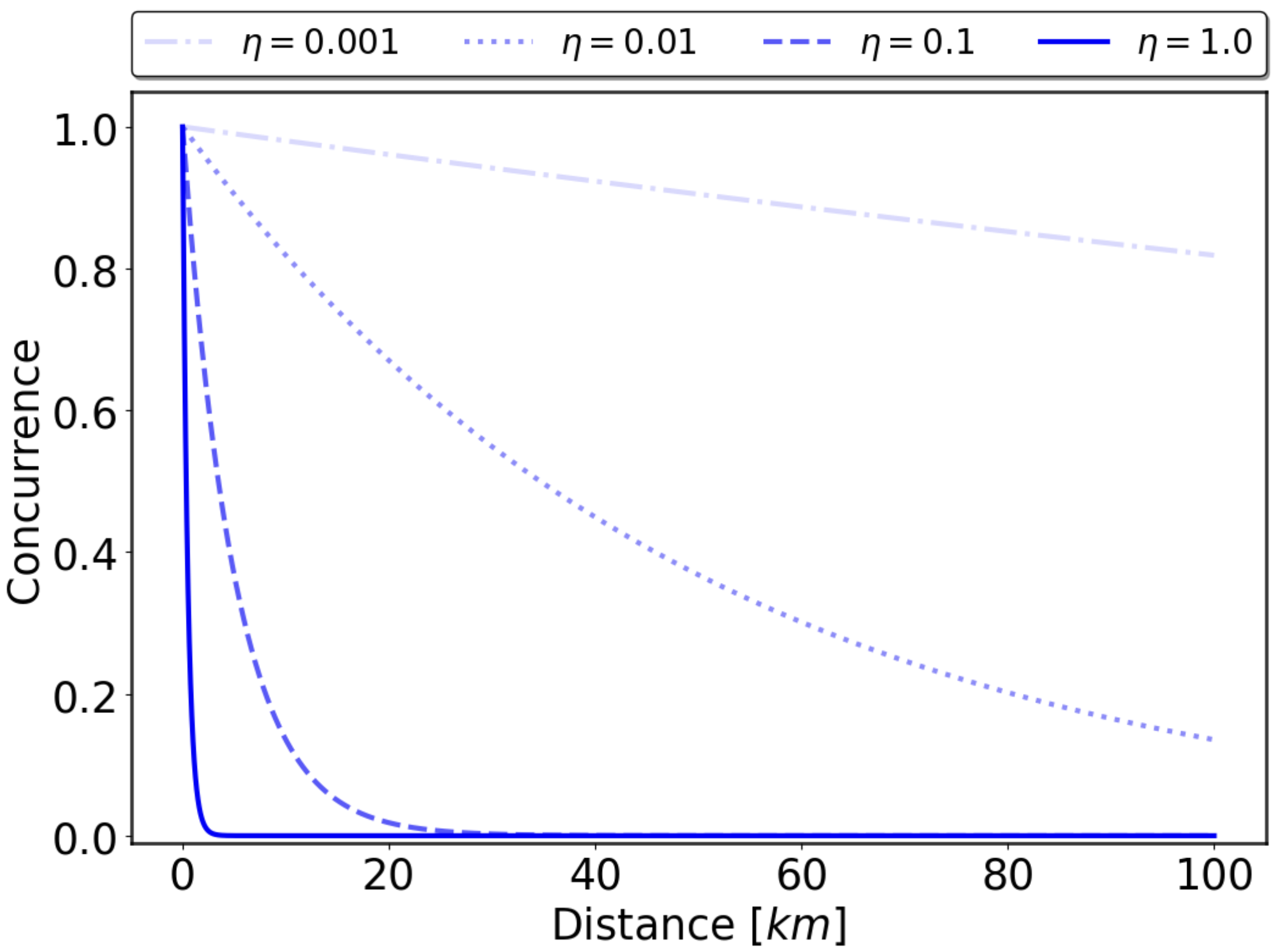}
    \renewcommand{\figurename}{Figure}
    \caption {Entanglement decay with propagation distance for a fixed correlation length $L_c = 10$ m and various coupling strengths. As expected, propagation distance is inversely proportional to coupling strength $\eta$.}
    \label{fig:concur_vs_distance_noDD_lambdas}
\end{figure}

Fig~\ref{fig:concur_vs_distance_noDD_gammas} shows numerical results for the entanglement, as measured by concurrence, as a function of propagation distance for several fiber correlation lengths and with $\eta=0.1$. In this case, the dynamics shifts from Markovian to non-Markovian as the correlation length and consequently, bath memory time increases. In the long correlation length regime, the concurrence tends to dissipate faster and exhibits the signature entanglement revival indicative of significant bath memory effects. As the correlation length decreases, the birefringence noise becomes uncorrelated, and the concurrence decays relatively slowly and without revivals, indicating Markovian dynamics.
Fig~\ref{fig:concur_vs_distance_noDD_lambdas} shows entanglement decay for varying system-bath coupling strengths at a constant bath memory time. Expectedly, the concurrence decays at the rate proportional to coupling strength.
\subsection*{Quantifying non-Markovianity}
\begin{figure}[h]
    \centering
    \includegraphics[width=\columnwidth]{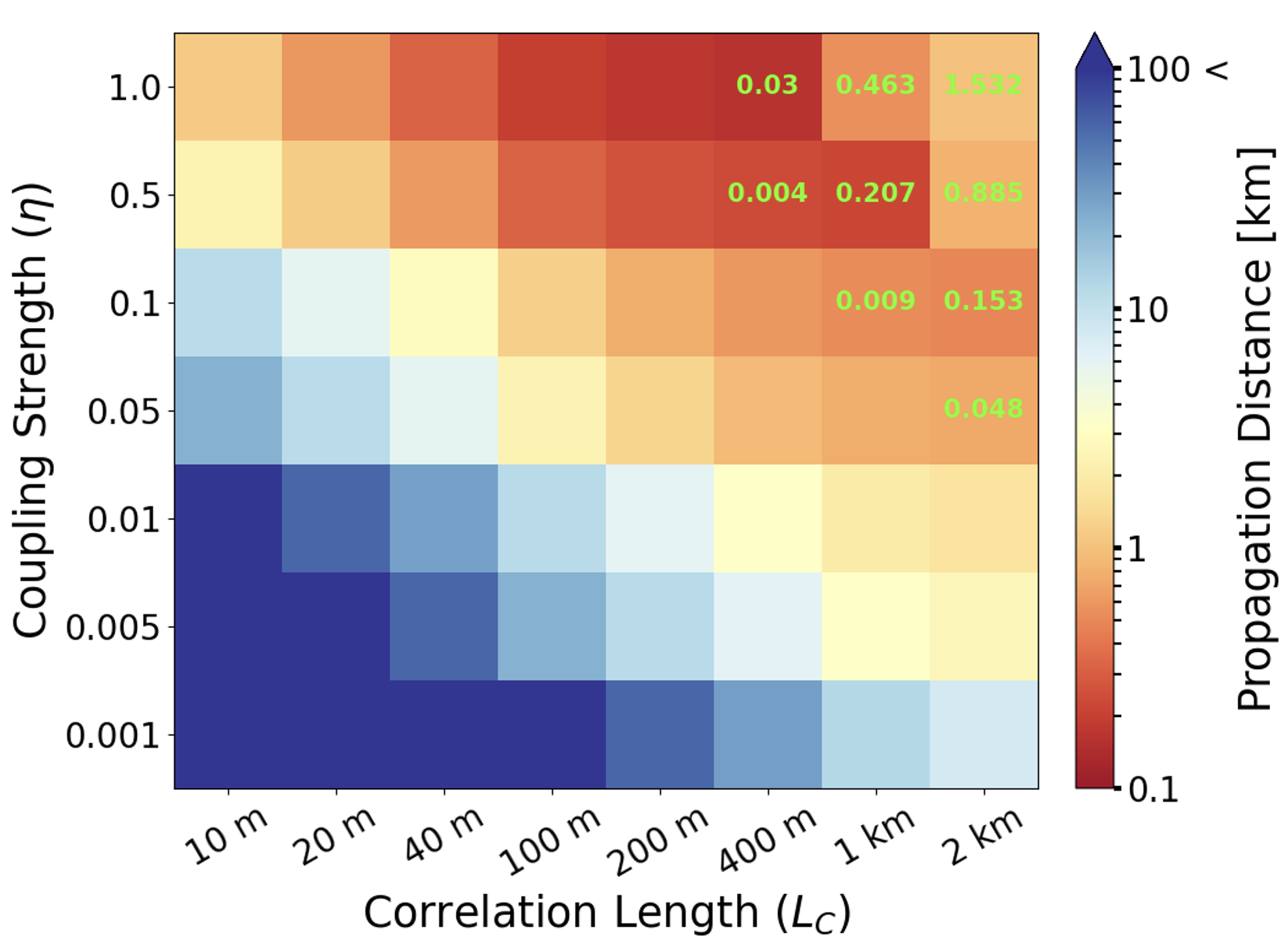}
    \renewcommand{\figurename}{Figure}
    \caption {Distance propagated before the concurrence decays to 0.1 for different combinations of coupling strengths $\eta$'s and correlation lengths $L_c$. Note that the colorbar indicates the distance in km and is in log scale. Values of $\mathcal{N}$, displayed in green on the 2D plot, represent the non-Markovian nature of the dynamics for the combinations of $\eta$ and $L_c$ values.}
     \label{fig:2d_distance_nm}
\end{figure}

The dynamics of entanglement may display characteristics of non-Markovian behavior when the coupling strength is substantial or the correlation length of the fiber is extended. A primary feature of such dynamics is the revival of entanglement due to the memory effects of the bath. Using these revivals, non-Markovianity can be quantified using the measure~\cite{rivas2010entanglement, gaikwad2024entanglement}:
\begin{equation}
\label{eq:nonmark}
    \mathcal{N} = \int_{t_0}^{t_f} dt \left | \frac{d\mathcal{E}(\rho_{SB}(t))}{dt} \right | - \Delta \mathcal{E},
\end{equation}
with $\mathcal{E}$ is the amount of entanglement and $\rho_{SB}$ is the system-bath density matrix. In this work, amount entanglement $\mathcal{E}$ is quantified by concurrence. To determine $\mathcal{N}$ in our simulations, we numerically calculate the discrete-time derivative of the concurrence and integrate it over timesteps from initial time $t_0$ to final time $t_f$ and $\Delta \mathcal{E} = \mathcal{E}(t=t_0) - \mathcal{E}(t=t_f) $. While $\mathcal{N} = 0$ in the case of Markovian dynamics, entanglement revivals contribute positively towards time derivative and make $\mathcal{N} > 0$ in the case of non-Markovian dynamics. Numerical values of $\mathcal{N} > 0$ are displayed in green on Fig~\ref{fig:2d_distance_nm} and appear on the bottom left of the figure where both $\eta$ and $L_{c}$ are larger. It predicts that fibers with longer correlation lengths and large fluctuations in birefringence could show non-Markovian character. Value of $\mathcal{N}$ depends more critically on correlation length and decays exonentially as it decreases. 

Transition from non-Markovian to Markovian dynamics and corresponding increase in propagation distance can be seen in Fig.~\ref{fig:2d_distance_nm} with transition region across the diagonal.

\section{Dynamical Decoupling}
\label{sec:DD}
Practical implementations of quantum technologies such as long-distance quantum communication and distributed quantum sensing would need entanglement distribution over hundreds of kilometers via optical fiber. Decoherence effects caused by random fluctuations in the birefringence of these fibers could severely limit such long-distance entanglement distribution unless active measures compensate for it. In this section, we evaluate the effectiveness of dynamical decoupling pulses \cite{Viola} to preserve entanglement in optical fibers for long-distance entanglement distribution.

Dynamical decoupling is a technique used in quantum control to mitigate the effects of environmental noise on qubits by applying a series of carefully timed control pulses. In the context of photonic polarization qubits, these pulses can be implemented by inserting half waveplates in the path of photons.
A suitably chosen DD pulse sequence with the interval between consecutive pulses shorter than the noise correlation time can ``average out" the environmental noise effects before they cause significant decoherence. The simplest choice of the pulse sequence that eliminates system-environment interaction to first order is known as the Carr-Purcell-Meiboom-Gill (CPMG) DD pulse sequence \cite{CPMG,PhysRevLett.82.2417}. It is an equidistant two-pulse sequence, ubiquitously used for decoupling of the system from the environment. Examples of more complex sequences to cancel higher-order errors are concatenated (CDD) and Uhrig (UDD) dynamical decoupling pulses.
\begin{figure}[h]
    \centering
    \includegraphics[width=0.9\columnwidth]{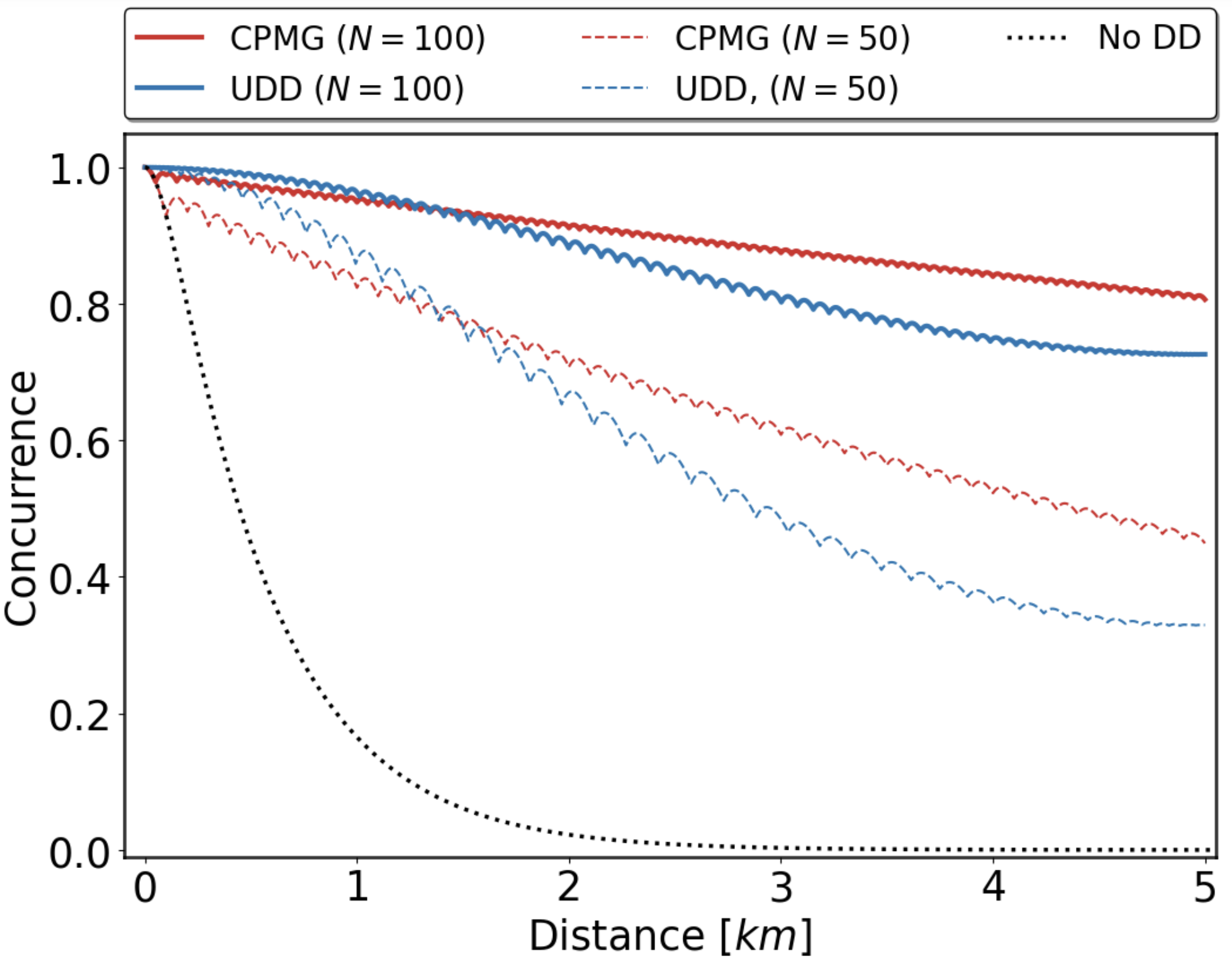}
    \renewcommand{\figurename}{Figure}
    \caption {Evolution of the concurrence over a propagation distance of 5 km with and without correcting half waveplates. CPMG and UDD sequences here contains 100 half waveplates in each and $\eta = 0.1$ and $L_c = 100 m$. Dotted curve represents entanglement decay without any correcting waveplates.}
    \label{fig:dd_comparison-3curves}
\end{figure}

We propose countering the birefringence-induced disentangling of photonic qubits by realizing a DD pulse sequence by inserting half-wave plates, which act as $\pi$ pulses, at predetermined positions in the optical fiber. In the HEOM formalism, we include the DD pulses in terms of a time-dependent Hamiltonian so that the system-bath evolves under the action of the total Hamiltonian:
\begin{equation}
\begin{split}
    H =& H_S+ H_B + H_{SB} + H_C(t)\\
    &H_C(t) = \sum_{n=1}^{N} h(t)  \hat{\sigma}^x_1 \otimes \hat{\sigma}^x_2
\end{split}
\end{equation}
with control Hamiltonian $H_C(t)$, which is a series of N $\pi$ pulses around the X-axis of the Bloch sphere and with the time-dependent amplitudes $h(t)$. In this work we consider two such schemes which differ from each other in terms of the spacing between two consecutive pulses. For a total evolution time $T$, the CPMG sequence uses $N$ uniformly spaced $\pi$ pulses, while in the UDD sequence, the spacing between two consecutive pulses is non-uniform. Any $j$th pulse is applied at times $t_j$ given as \cite{uhrig2007keeping}:
\begin{equation}
t_j = T \sin^2\left ( \frac{j\pi}{2(N+1)} \right ).
\end{equation}
\begin{figure*}[ht]
    \centering
    \includegraphics[width=\textwidth]{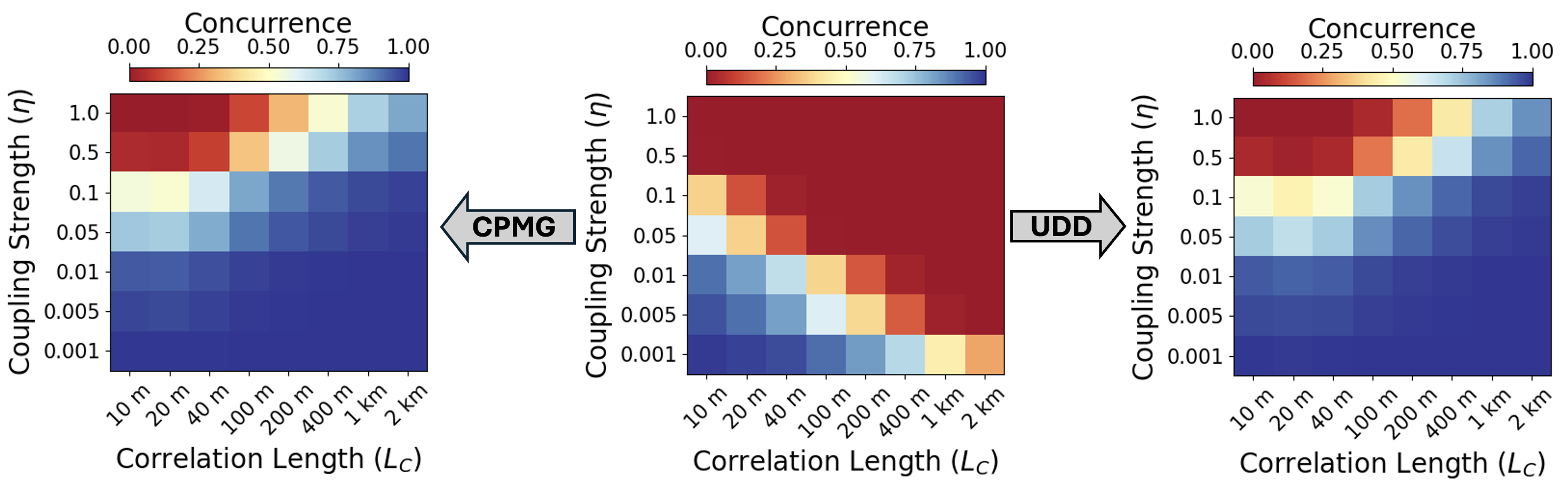}
    \renewcommand{\figurename}{Figure}
    \caption {Comparison of effectiveness of CPMG and UDD pulse sequence for total of 100 half waveplates in each. Central 2D plot is of the concurrence at the end of 5 km propagation distance without any half waveplates.}
    \label{fig:2d_cpmg_udd_comparison}
\end{figure*}
Fig.~\ref{fig:dd_comparison-3curves} shows entanglement evolution in the optical fiber over the length of 5 km of propagation distance under the action of $N=100$ and $N=50$ half waveplates in both CPMG and UDD configuration. Value of the birefringence is assumed to fluctuate by $10\%$ around the mean value $\overbar{\Delta n}$ and correlation length $L_c$ is assumed to be 100 m. $\pi$ pulses are assumed to be rectangular in shape with finite values of pulse interval $\tau_p$ and amplitude $A_p$ satisfying the relation $\tau_p * A_p = \pi/2$. For the $N=100$ and $N=50$ CPMG sequence used here, the distance between two consecutive half-waveplates is 50 and 100 meters, respectively. Without correcting half waveplates, the entanglement decays to zero over 2-3 km. However, even a few half waveplates per correlation length can significantly protect entanglement, and the level of residual concurrence increases with the number of waveplates in the DD sequence. In UDD sequence separation between initial waveplates is much smaller and it increases in the middle and gets smaller again in the end part of the sequence. This is reflected in both the cases $N=100$ and $N=50$ in Fig.~\ref{fig:dd_comparison-3curves} where the UDD sequence initially performs better than the CPMG sequence but cross-over occurs in the middle of the evolution as the waveplate separation increases, resulting in poor protection against entanglement decay. Rate of this decay slows down in the end when waveplate separation decreases again. 

\subsection*{Effectiveness of CPMG and UDD}
Performance comparison of different DD sequences is an active area of research \cite{ezzell2023dynamical, gong2020noise, niu2022effects}. Although more complex and elaborate DD sequences such as periodic (PDD), concatenated (CDD), nested (NUDD), and quadratic (QDD) have been developed to increase the decoupling order, we focus on traditional sequences such as CPMG which have been shown to perform equally well in case of large pulse separation \cite{ezzell2023dynamical}. 
We evaluate the effectiveness of the CPMG and UDD sequence by comparing the amount of entanglement protected after 5 km of propagation through optical fiber of varying properties. 

Fig.~\ref{fig:2d_cpmg_udd_comparison} (center) shows the residual entanglement, as measured by the concurrence, of the input Bell state after 5 km of propagation through the fiber with the corresponding $\eta$ and $L_c$ and without any correcting DD pulses. Except for upper right corner, which corresponds to very low fluctuations in birefringence magnitude and fibers with short correlation length, entanglement either disappears or decays significantly. 
The left and right plots of Fig.~\ref{fig:2d_cpmg_udd_comparison} are of protected entanglements under the same conditions using the CPMG and UDD sequences, respectively. Each sequence uses a total of 100 half waveplates along the total length of 5 km of the fiber. 
\begin{figure}[h]
    \centering
    \includegraphics[width=\columnwidth]{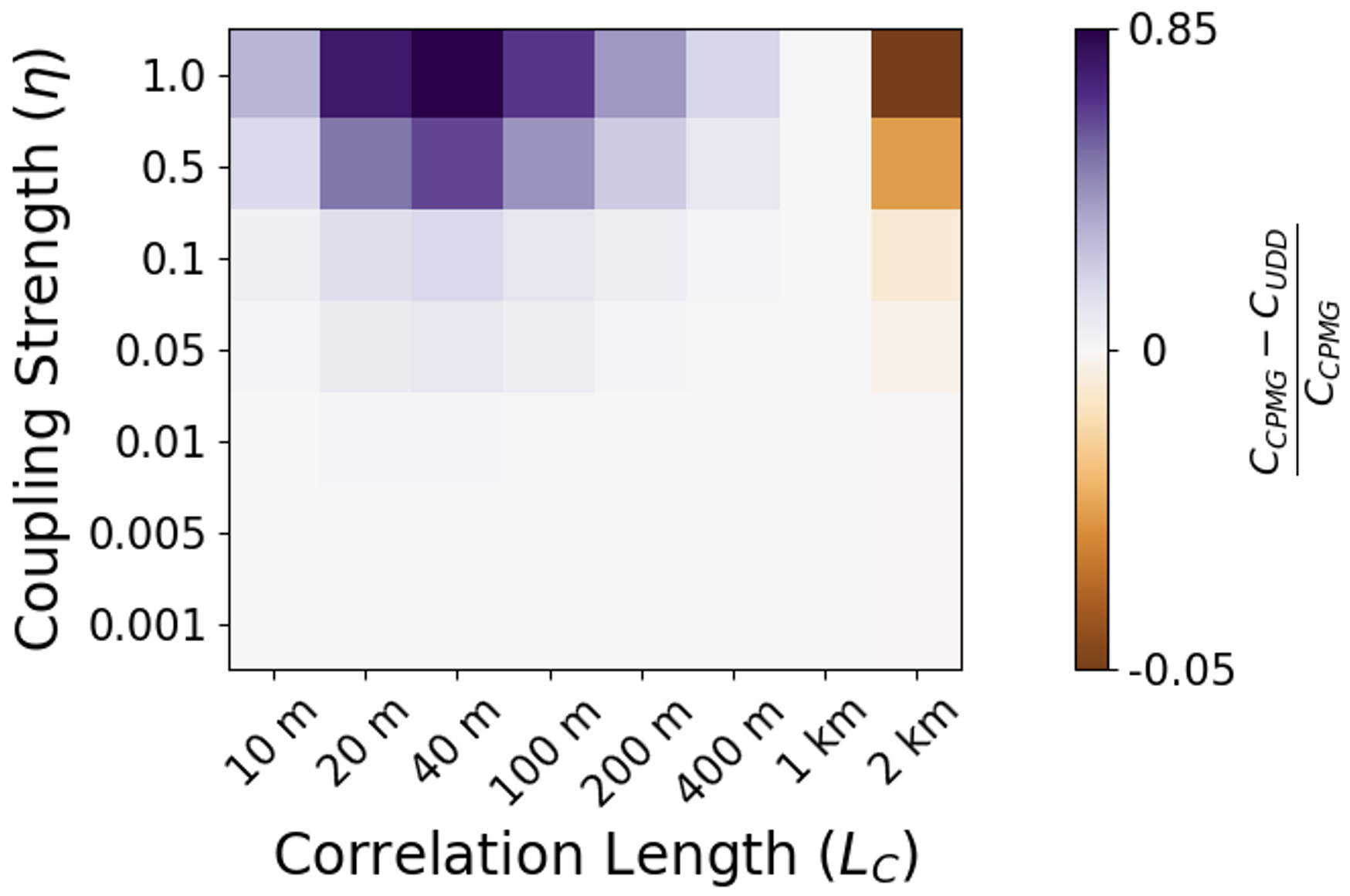}
    \renewcommand{\figurename}{Figure}
    \caption {Comparison of effectiveness of CPMG and UDD pulse sequence in terms of difference in the residual concurrence after 5 km of propagation distance. Total of 100 half waveplates in each. }
    \label{fig:diff-dd-2d}
\end{figure}
For majority of the parameter space considered, both DD sequences manage to maintain the concurrence in the $0.85-1.0$ range and this performance improves as the total number of pulses increases. Fibers with short correlation length and high variation in birefringence, which belong to the upper left corner of Fig.~\ref{fig:2d_cpmg_udd_comparison}, remain immune to the effects of DD pulses. This can be explained by noting that DD schemes work under the assumption that correcting DD pulses are applied before the noise changes from one to appreciably different another realization. For pulses separated by a distance greater than the fiber correlation length, DD is gradually rendered ineffective for large fluctuations in mean birefringence.

Fig.\ref{fig:diff-dd-2d} shows the performance comparison of the CPMG and UDD sequences, quantified by the difference between the residual concurrences as $\frac{C_{CPMG}-C_{UDD}}{C_{CPMG}}$ over the range of parameter values. Across the majority of the parameter space, both DD sequences exhibit comparable performance. The regions shaded in purple and orange indicate areas where the CPMG and UDD sequences, respectively, demonstrate a marginal advantage.

\section{Conclusion}
\label{sec:IV}
To summarize, we theoretically analyze the birefringence-induced decay of polarization-entanglement in optical fibers using principles of open quantum system. We use the HEOM formalism that, unlike the commonly used Lindblad master equation, is not limited to Markovian dynamics and hence allows us to study the impact of birefringence noise on entanglement dynamics in both the Markovian and non-Markovian regime. For a range of realistic values of parameters such as birefringence and correlation length, we model entanglement distribution via fibers and obtain estimates of limits imposed on it by birefringence-induced decoherence. We show that a combination of large fluctuations in birefringence and long correlation lengths can lead to non-Markovian dynamics, the sudden death of entanglement, and characteristic revivals at later times. We use a revival-based measure to quantify the non-Markovianity across the parameter range.

To counter the entanglement decay due to birefringence noise we proposed realizing dynamical decoupling in optical fibers through the insertion of half waveplates. To study our approach we included a time-dependent Hamiltonian in our HEOM model and studied timescales smaller than the bath correlation time to capture the effect of these pulses on system-bath dynamics. Specifically, we used the CPMG and Uhrig DD schemes and showed that they can reverse the undesired phase errors introduced by random fluctuations in fiber birefringence and thereby extend the range of polarization-entanglement distribution using optical fibers. We also evaluated and compared the performance of these two DD schemes and showed that the conceptually straightforward CPMG approach performed either equally well or marginally better in the relevant parameter ranges.

\section*{Acknowledgement}
The authors thank Stav Haldar for helpful discussions. This work was supported by the Army Research Office Multidisciplinary University Research Initiative (ARO MURI) through grant number W911NF2120214. PB and HL also acknowledge the support of the US Air Force Office of Scientific Research and the US-Israel Binational Science
Foundation.

\bibliography{References.bib}
\bibliographystyle{unsrt}

\end{document}

%% file: preamble.tex
\usepackage{amsthm}
\usepackage{mathtools}
\usepackage{physics}
\usepackage{xcolor}
\usepackage{graphicx}
\usepackage[left=23mm,right=13mm,top=35mm,columnsep=15pt]{geometry} 
\usepackage{adjustbox}
\usepackage{placeins}
\usepackage[T1]{fontenc}
\usepackage{lipsum}
\usepackage{csquotes}